# Prerequisite of superconductivity: SDW rather than tetragonal structure in double-layer $La_3Ni_2O_{7-\delta}$

Mengzhu Shi[1,2,†], Di Peng[6,†], Yikang Li[1,2], Zhenfang Xing[7], Yuzhu Wang[3], Kaibao Fan[1,2], Houpu Li[1,2], Rongqi Wu[1,2], Zhidan Zeng[7], Qiaoshi Zeng[6,7], Jianjun Ying[1,5], Tao Wu[1,2,4,5,*], and Xianhui Chen[1,2,4,5,*]

1. CAS Key Laboratory of Strongly-coupled Quantum Matter Physics, Department of Physics, University of Science and Technology of China, Hefei, Anhui 230026, China

2. Hefei National Laboratory for Physical Sciences at the Microscale, University of Science and Technology of China, Hefei, Anhui 230026, China

3. Shanghai Synchrotron Radiation Facility, Shanghai Advanced Research Institute, Chinese Academy of Sciences, Shanghai, China

4. Collaborative Innovation Center of Advanced Microstructures, Nanjing University, Nanjing 210093, China

5. Hefei National Laboratory, University of Science and Technology of China, Hefei 230088, China

6. Shanghai Key Laboratory of Material Frontiers Research in Extreme Environments (MFree), Institute for Shanghai Advanced Research in Physical Sciences (SHARPS), Shanghai, China

7. Center for High Pressure Science and Technology Advanced Research, Shanghai, China

†These authors contributed equally to this work
*Correspondence to: wutao@ustc.edu.cn, chenxh@ustc.edu.cn,

**The pressure-induced high-temperature ($T_c$) superconductivity in nickelates $La_3Ni_2O_{7-\delta}$ has sparked significant interest to explore its superconductivity at ambient pressure. $La_{n+1}Ni_nO_{3n+1}$ (n=2, 3) adopts an orthorhombic structure with tilted $NiO_6$ octahedra and undergoes a spin density-wave (SDW) transition at ambient pressure, while the octahedral tilting and the SDW are suppressed by pressure, and high pressure induces a structural transition from orthorhombic to tetragonal, and the high-Tc superconductivity is achieved in the tetragonal structure. This tetragonal structure is widely believed to be crucial for the pressure-induced** **superconductivity. Whether the pressure-stabilized tetragonal structure is a prerequisite for achieving nickelate superconductivity at ambient pressure is under hot debate. Here, by post-annealing of the orthorhombic $La_3Ni_2O_{7-\delta}$ as grown microcrystals with noticeable oxygen defects in high oxygen pressure environment, tetragonal $La_3Ni_2O_{6.96}$ single crystals are successfully obtained at ambient pressure. In contrast to the orthorhombic $La_3Ni_2O_{7-\delta}$, the tetragonal $La_3Ni_2O_{7-\delta}$ exhibits metallic behavior without a SDW transition at ambient pressure. Moreover, no superconductivity is observed at high pressure up to ~ 70 GPa. On the other hand, by utilizing Helium as the pressure medium, we have revisited the superconducting structure in pressurized orthorhombic $La_3Ni_2O_{6.93}$. Our results indicate that the orthorhombic structure is quite robust against pressure, and no structural transition from orthorhombic to tetragonal happens, and the superconductivity under high pressure is achieved in orthorhombic structure rather than tetragonal structure claimed previously. All these results suggest that tetragonal structure is not prerequisite for achieving superconductivity in $La_3Ni_2O_{7-\delta}$. In addition, the absence of both density-wave transition and pressure-induced superconductivity in tetragonal $La_3Ni_2O_{6.96}$ suggests that the underlying physics of SDW order and superconductivity has a deep correlation. It should be emphasized that the most important finding is that the structure and SDW order in $La_3Ni_2O_{7-\delta}$ strongly depend on the oxygen concentration at ambient pressure. Our present work imposes stringent constraints on the underlying mechanism for pressure-induced superconductivity in nickelates and sheds new light on the exploration of nickelate superconductivity at ambient pressure.**

Search for high-temperature superconductivity in nickelates has attracted a long-term interest in community after the discovery of cuprate superconductivity[1-7]. In 2019, the breakthrough in infinite-layer nickelate $Nd_{1-x}Sr_xNiO_2$ thin film paved a new wave in this field[8]. Recently, the nickelate superconductor family has been successfully expanded to the Ruddlesden-Popper (RP) phases $La_{n+1}Ni_nO_{3n+1}$, with n = 2[9] and n = 3[10-13]. The superconducting transition temperature ($T_c$) in $La_3Ni_2O_{7-\delta}$ exceeds the liquid-nitrogen boiling temperature, suggesting a new family of high-temperature superconductors[14-26]. However, unlike the infinite-layer nickelate thin films, the realization of superconductivity in these RP nickelates requires extremely high-pressure conditions, which hinders most spectroscopic measurements of the superconducting state and complicates the study of the underlying mechanism[14]. Exploration of superconductivity at ambient pressure in the RP phase $La_{n+1}Ni_nO_{3n+1}$ not only offers a promising strategy to overcome the above challenges, but also has significant importance for the application of nickelate superconductors.

In $La_3Ni_2O_{7-\delta}$, the $NiO_6$ octahedra are tilted away from the longest axis at ambient pressure, resulting in an orthorhombic structure with space group of *Amam* (Fig. 1a) rather than tetragonal structure with space group of *I4/mmm* (Fig. 1b). In the double-layer $La_3Ni_2O_{7-\delta}$ with *Amam* structure, the bond angle of Ni-O-Ni between adjacent $NiO_6$ octahedra is 168 degrees. Earlier transport and spectroscopic measurements have revealed a density-wave transition around 150 K at ambient pressure[27-33]. A similar density-wave transition has also been observed in the trilayer nickelate $La_4Ni_3O_{10-\delta}$, in which an intertwined density wave with both charge and spin order is revealed (~ 136 K)[38]. As pressure increases, the unit cell volume shrinks significantly, making the tilted $NiO_6$ octahedra unstable. Recent X-ray diffraction (XRD) experiments under high pressure revealed that high pressure induces a structural transition from orthorhombic to tetragonal, and the pressure-induced superconductivity occurs in a tetragonal structure with *I4/mmm*[34-36]. In *I4/mmm* structure, the bond angle of Ni-O-Ni between adjacent $NiO_6$ octahedra becomes 180 degrees. In earlier theory, this change of the bond angle of Ni-O-Ni can significantly affect the interlayer coupling between NiO planes and is thought to be important for achieving the superconductivity under high pressure[9]. Therefore, to stabilize the tetragonal structure becomes a prerequisite for exploring nickelate superconductivity at ambient pressure. On the other hand, recent density functional theory (DFT) calculation also indicates that this change of the bond angle of Ni-O-Ni has no significant effect on the band structure, especially for $d_z^2$ orbital dominant band[41]. Following this line, the tetragonal structure is not crucial for superconductivity. Instead, the suppression of density wave transition in orthorhombic double-layer $La_3Ni_2O_{7-\delta}$ is important for achieving superconductivity at ambient pressure. In the present study, we successfully obtain the tetragonal structure with *I4/mmm* in $La_3Ni_2O_{7-\delta}$ single crystals at ambient pressure. Moreover, the superconducting structure in pressurized orthorhombic

$La_3Ni_2O_{7-\delta}$ has been revisited. Our results indicate that tetragonal structure is not a necessary condition for achieving superconductivity in $La_3Ni_2O_{7-\delta}$. Existence of spin-density-wave (SDW) order at ambient pressure seems more vital for pressure-induced superconductivity.

**Tetragonal $La_3Ni_2O_{7-\delta}$ at ambient pressure**

The as-grown $La_3Ni_2O_{7-\delta}$ samples are obtained by utilizing the melt salt as the flux (see Methods for details), which are crystallized into an *Amam* orthorhombic structure with noticeable oxygen defects and exhibit an insulating behavior (Extended Data Fig. 1). The recent multislice electron ptychography measurement indicates that the oxygen deficiency appears mainly at the inner apical oxygen sites and such a kind of oxygen defect makes the oxygen content in $La_3Ni_2O_{7-\delta}$ single crystal quite inhomogeneous[17]. After an additional post annealing process under oxygen pressure of 10-15 bar to repair the oxygen defects, the temperature-dependent resistance becomes metallic while the orthorhombic structure is kept in $La_3Ni_2O_{7-\delta}$ samples (Fig. 1d and Extended Data Fig. 1). Pressure-induced high-temperature superconductivity can be observed in these annealed orthorhombic $La_3Ni_2O_{7-\delta}$ samples, which will be carefully discussed later. At this stage, the oxygen content in orthorhombic $La_3Ni_2O_{7-\delta}$ is determined to be 6.93 by the XRD refinement on the selected microcrystal (Table 1). With further increasing the annealing oxygen pressure to 150 bar, the orthorhombic structure is not stable anymore and an *I4/mmm* tetragonal structure becomes stable in the $La_3Ni_2O_{7-\delta}$ single crystal. As shown in Fig. 1c and 1d, the powder XRD pattern of the tetragonal $La_3Ni_2O_{7-\delta}$, obtained by grinding several microcrystal pieces, can be well indexed by the calculated diffraction peaks for a *I4/mmm* tetragonal structure. As shown in Fig. 1d, the two peaks are well indexed as (105) and (110) in the tetragonal structure. No additional splitting due to orthorhombic distortion is observed in this range of 2θ. In comparison, obvious peak splitting can be observed for the orthorhombic $La_3Ni_2O_{7-\delta}$ as shown in Fig. 1d. In addition, the oxygen content in the tetragonal $La_3Ni_2O_{7-\delta}$ at ambient pressure is also determined by the XRD refinement on the selected microcrystal, and is about 6.96 slightly higher than that in orthorhombic $La_3Ni_2O_{6.93}$. In tetragonal $La_3Ni_2O_{6.96}$, the bond angles of Ni-O-Ni between adjacent $NiO_6$ octahedrons along the longest axis change from 168° to 180°.

Moreover, the in-plane Ni-O-Ni bonding angle at ambient pressure is 178.9° which is quite consistent with the previously reported value of 178.4° for tetragonal $La_3Ni_2O_{7-\delta}$ under high pressure[34]. More detailed crystal data and refinement information are presented in Table 1.

It should be noted that no density-wave and no superconducting transition is observed in the temperature-dependent resistance of tetragonal $La_3Ni_2O_{7-\delta}$ at ambient pressure (Extended Data Fig. 1). This is one of the interesting findings in the present work. Since the signature of density-wave transition is weak, and even no feature can be seen in resistance in orthorhombic $La_3Ni_2O_{7-\delta}$, we choose to utilize magnetic torque technique to detect the possible density-wave transition in tetragonal $La_3Ni_2O_{7-\delta}$ at ambient pressure. First, we measure the magnetic torque in orthorhombic $La_3Ni_2O_{7-\delta}$ to make a benchmark for SDW transition. As shown in Fig. 1e, the temperature-dependent magnetic torque $\tau(T)$ of the orthorhombic $La_3Ni_2O_{7-\delta}$ shows a clear *S*-shape behavior around 150 K, which is ascribed to the reported SDW transition by different techniques[29,31,32]. This SDW transition is more clearly seen in the $d\tau/dT$ vs $T$ curves (Fig. 1f). Furthermore, as shown in Fig. 1g, the temperature-dependent $\tau(T)$ is also measured in the tetragonal $La_3Ni_2O_{7-\delta}$. In contrast to orthorhombic $La_3Ni_2O_{7-\delta}$, no SDW transition can be resolved in our present measurement. This result indicates that SDW transition is absent in the tetragonal $La_3Ni_2O_{7-\delta}$. More implications of these results will be discussed later. Next, we will further explore the pressure-induced superconductivity in both orthorhombic and tetragonal $La_3Ni_2O_{7-\delta}$ samples.

**Transport properties of pressurized orthorhombic and tetragonal $La_3Ni_2O_{7-\delta}$**

To investigate the pressure-induced superconductivity in $La_3Ni_2O_{7-\delta}$ with different structures, we performed resistance measurements on selected microcrystals under high pressures. Helium gas was loaded as the pressure transmitting medium to provide homogeneous pressure environment[39]. The transport properties of the orthorhombic $La_3Ni_2O_{7-\delta}$ microcrystal under various pressures are shown in Figure 2. In pressurized orthorhombic $La_3Ni_2O_{7-\delta}$, the room-temperature resistance gradually decreases with increasing the pressure. At 15.3 GPa, the pressure-induced superconductivity appears in the temperature-dependent resistance but the superconducting transition is still broad. By further increasing pressure, the superconducting transition becomes more pronounced and a sharp superconducting transition can be achieved above 20 GPa as shown in Fig. 2a and 2b. The highest onset temperature of the superconducting transition reaches to about 80 K while zero resistance temperature is about 43.5 K to 45 K for two samples as shown in the inset of Fig. 2c and 2d. These results on pressure-induced superconductivity are even better than the best superconducting transition in pressurized $La_3Ni_2O_{7-\delta}$ in reported literatures[9,16], indicating the importance of the homogeneous pressure environment for the high-pressure measurements and the high quality of our annealed orthorhombic $La_3Ni_2O_{7-\delta}$ samples. By increasing the pressure above 20 GPa, the superconducting transition temperature starts to decrease slowly with increasing the pressure which is also consistent with previous high-pressure measurements[9,16]. The nature of optimal superconductivity is also examined by applying a magnetic field perpendicular to the Ni-O planes. As shown in Fig. 2c and 2d, the superconducting transition under different magnetic fields is measured in two superconducting orthorhombic $La_3Ni_2O_{7-\delta}$ samples. The field-dependent superconducting transition temperatures are fitted by a Ginzburg–Landau (GL) model (Extended Data Fig. 2) and the extracted upper critical fields are comparable with the reported values in literatures[9,16].

In the case of tetragonal $La_3Ni_2O_{7-\delta}$, we first try to study the pressure effect on resistance by utilizing Helium gas as the pressure transmitting medium. As shown in Fig. 3a and 3b, the temperature dependent resistance shows the metallic behavior in the whole temperature range, and no significant pressure effect on the overall metallic behavior is observed with increasing pressure. Although the room-temperature resistance continuously decreases with increasing pressure, no superconductivity is observed up to 30.3 GPa. Then, we try to extend the study of pressure effect to higher pressure by utilizing NaCl as the pressure transmitting medium. As shown in Fig. 3c and 3d, superconductivity is still absent up to 68.2 GPa. The fact that the metallic behavior without density-wave transition persists to high pressures indicates a robust ground state in tetragonal $La_3Ni_2O_{7-\delta}$, which is distinct from the pressurized orthorhombic $La_3Ni_2O_{7-\delta}$. On the other hand, the absence of superconductivity in pressurized tetragonal $La_3Ni_2O_{7-\delta}$ also poses a great challenge to the idea that the tetragonal structure is crucial for pressure-induced superconductivity. Next, we further study the high-pressure structure in orthorhombic and tetragonal $La_3Ni_2O_{7-\delta}$ to check the pressure-induced structural transition.

**Structural evolution with pressure for the orthorhombic and tetragonal $La_3Ni_2O_{7-\delta}$**

We conducted high-pressure powder XRD measurements on both orthorhombic and tetragonal $La_3Ni_2O_{7-\delta}$ samples by utilizing Helium gas as the pressure transmitting medium, as illustrated in Fig. 4 and Extended Data Figs. 3 to 6. In principle, the Helium pressure transmitting medium offers a more homogeneous pressure environment compared to the previous measurements by utilizing liquid silicon oil as transmitting medium[34-36]. In pressurized tetragonal $La_3Ni_2O_{7-\delta}$, the powder XRD data under different pressures are well fitted by a tetragonal structure with the space group of *I4/mmm* (see examples for 2.5 GPa and 22.1 GPa in Fig. 4b and Extended Data Fig. 3), and no peak splitting related to the orthorhombic distortion is resolved. As shown in Fig. 4c, the determined lattice parameters (*a*-axis and *c*-axis) and cell volume exhibit an almost linear decrease with increasing pressure, suggesting the absence of structural transition in tetragonal $La_3Ni_2O_{7-\delta}$ under pressure up to 26 GPa.

Similarly, our present work found that the orthorhombic $La_3Ni_2O_{7-\delta}$ also exhibits a stable orthorhombic structure up to 25.6 GPa as indicated by the powder XRD data (Fig. 4d and 4e). This result is quite different from the previous observation in orthorhombic $La_3Ni_2O_{7-\delta}$[34-36]. In details, the splitting of the (020) and (002) peaks, attributed to orthorhombic distortion, is observed at all pressures (see examples for 1.5 GPa and 21.8 GPa in Fig. 4e and Extended Data Figs. 4 and 6). The derived lattice parameters and cell volume for orthorhombic $La_3Ni_2O_{7-\delta}$ are illustrated in Fig. 4f. It is obvious that the lattice parameters exhibit an almost linear decrease with pressure, suggesting the absence of structural transition under pressure. As shown in Extended Data Fig. 4, the powder XRD patterns for $La_3Ni_2O_{7-\delta}$ at 1.5 GPa and 21.8 GPa can be accurately fitted by *Amam* space group. All these results strongly support that the pressure-induced superconducting phase remains in the orthorhombic structure. To compare with the reported literatures, we conducted pressure-dependent powder XRD measurements on orthorhombic $La_3Ni_2O_{7-\delta}$ by utilizing silicon oil as the pressure transmitting medium which is a method commonly employed in prior studies[34-36]. As shown in Extended Data

Fig. 7, the evolution of diffraction peaks (020) and (002) is quite consistent with previous reports and looks like gradually merging under pressure above 11 GPa. Considering the present results with Helium gas as the pressure transmitting medium, this phenomenon is probably due to the inhomogeneity of the pressure environment, underscoring the importance of pressure homogeneity in high-pressure experiments. These results definitely indicate no pressure-induced structural transition in the orthorhombic $La_3Ni_2O_{7-\delta}$, and the structural transition reported previously could be ascribed to pressure inhomogeneity which leads to the broadening of the diffraction peaks and makes the (020) and (002) diffraction peaks to merge together consequently.

Now we discuss the potential implications of the present work for exploring the RP-phase nickelate superconductors at ambient pressure. Previous studies have proposed that one of the key roles of high pressure induced superconductivity lies in stabilizing the tetragonal structure without octahedral tilting, as observed in both $La_3Ni_2O_{7-\delta}$ and $La_4Ni_3O_{10-\delta}$[11,34]. In fact, the role of $d_z^2$ orbital on the high-pressure superconductivity in $La_3Ni_2O_{7-\delta}$ and $La_4Ni_3O_{10-\delta}$ remains a subject of active debate[9,14]. Earlier theoretical proposals consider that the $3d_z^2$-derived bonding band below the Fermi energy will be lifted to form the γ pocket and plays a key role for the pressure-induced superconductivity while the octahedral tilting is suppressed with increasing pressure[9,14]. This theoretical idea highlights the significance of synthesizing tetragonal $La_3Ni_2O_{7-\delta}$ and $La_4Ni_3O_{10-\delta}$ crystals for achieving superconductivity at ambient pressure. However, such scenario is rather difficult to be proved due to the limiting of experimental characterizations at high pressure. Moreover, the recent DFT calculation suggests that the $3d_z^2$-derived bonding band might not cross the Fermi level and plays a less role on the pressure-induced superconductivity[41]. Therefore, the successful growth of the tetragonal $La_3Ni_2O_{7-\delta}$ at ambient pressure in this work is quite important to solve above argument.

It is well known that by significantly reducing the oxygen content, the tetragonal structure has been previously achieved at ambient pressure in $La_3Ni_2O_6$, where the apical oxygen atoms are completely removed[31]. However, in these oxygen-deficient tetragonal phases, no superconductivity has been reported and it shows an extremely insulating behavior[17,31]. This is significantly different from our tetragonal $La_3Ni_2O_{7-\delta}$ samples which are annealed under a high-pressure oxygen atmosphere. Although precise determination of the oxygen content remains challenging due to the small sample size, the refinements of single crystal X-ray diffraction (SC-XRD) data on the crystal structure suggest an oxygen content very close to stoichiometry (see Methods). In fact, the oxygen content of the tetragonal $La_3Ni_2O_{7-\delta}$ is found to be slightly higher than that of the orthorhombic $La_3Ni_2O_{7-\delta}$ by approximately 0.03 from the refinement of SC-XRD data on the crystal refinement. Consequently, the absence of superconductivity in the tetragonal $La_3Ni_2O_{7-\delta}$ samples cannot be attributed to the oxygen defects. This conclusion is further supported by the metallic resistivity behavior as shown in Fig. 3. Moreover, our high-pressure XRD measurements on the orthorhombic $La_3Ni_2O_{7-\delta}$ samples reveal that the structure of the high-pressure superconducting phase is orthorhombic rather than tetragonal, in stark contrast with previous studies[34]. It should be emphasized that the hydrostatic pressure is highly desired to detect such tiny differences in different structures under high pressures, which enlightening the importance of hydrostatic conditions in the high-pressure works. All these findings challenge the idea that high-pressure superconductivity emerges upon the formation of a tetragonal structure, and suggest that tetragonal structure is not a necessary condition for achieving superconductivity in double layer $La_3Ni_2O_{7-\delta}$. On the other hand, recent studies have confirmed SDW order in orthorhombic $La_3Ni_2O_{7-\delta}$ using various spectroscopic probes, including resonant inelastic X-ray scattering (RIXS) [40], nuclear magnetic resonance (NMR) [32], and muon spin rotation (μSR) experiments[29,37]. However, our transport and magnetic torque measurements reveal no evidence of any SDW transitions in the tetragonal $La_3Ni_2O_{7-\delta}$. Together with the absence of superconductivity, it implies that the SDW transition rather than the tetragonal symmetry as the primary condition for the pressure-induced superconductivity. Further experimental investigation on the tetragonal $La_3Ni_2O_{7-\delta}$ at ambient pressure is needed to understand the deep correlation between SDW order and superconductivity in double-layer $La_3Ni_2O_{7-\delta}$. Our finding that the structure type and SDW order in double layer $La_3Ni_2O_{7-\delta}$ strongly depend on the oxygen concentration paves the way to explore high-$T_c$ superconductivity at ambient pressure in nickelate superconductors.

**Reference**

1. Bednorz, J. G. & Müller, K. A. Possible high Tc superconductivity in the Ba−La−Cu−O system. *Z. Phys. B Condens. Matter* **64**, 189–193 (1986).

2. Norman, M. R. Materials design for new superconductors. *Rep. Prog. Phys.* **79**, 074502 (2016).

3. Azuma, M., Hiroi, Z., Takano, M., Bando, Y. & Takeda, Y. Superconductivity at 110 K in the infinite-layer compound $(Sr_{1-x}Ca_x)_{1-y}CuO_2$. *Nature* **356**, 775-776 (1992).

4. Anisimov, V. I., Bukhvalov, D. & Rice, T. M. Electronic structure of possible nickelate analogs to the cuprates. *Phys. Rev. B* **59**, 7901–7906 (1999).

5. Lee, K.-W. & Pickett, W. E. Infinite-layer $LaNiO_2$: $Ni^{1+}$ is not $Cu^{2+}$. *Phys. Rev. B* **70**, 165109 (2004).

6. Crespin, M., Levitz, P. & Gatineau, L. Reduced forms of $LaNiO_3$ perovskite. Part 1.— Evidence for new phases: $La_2Ni_2O_5$ and $LaNiO_2$. *J. Chem. Soc., Faraday Trans. 2* **79**, 1181–1194 (1983).

7. Hayward, M. A., Green, M. A., Rosseinsky, M. J. & Sloan, J. Sodium Hydride as a Powerful Reducing Agent for Topotactic Oxide Deintercalation: Synthesis and Characterization of the Nickel(I) Oxide $LaNiO_2$. *J. Am. Chem. Soc.* **121**, 8843–8854 (1999).

8. Li, D. et al. Superconductivity in an infinite-layer nickelate. *Nature* **572**, 624–627 (2019).

9. Sun, H. et al. Signatures of superconductivity near 80 K in a nickelate under high pressure. *Nature* **621**, 493–498 (2023).

10. Li, Q. et al. Signature of Superconductivity in Pressurized $La_4Ni_3O_{10}$. *Chin. Phys. Lett.* **41**, 017401 (2024).

11. Zhu, Y. et al. Superconductivity in tri-layer nickelate $La_4Ni_3O_{10}$ single crystals. *Nature* **631**, 531–536 (2024).

12. Zhang, M. et al. Superconductivity in tri-layer nickelate $La_4Ni_3O_{10}$ under pressure. Preprint at https://doi.org/10.48550/arXiv.2311.07423 (2024).

13. Sakakibara, H. et al. Theoretical analysis on the possibility of superconductivity in the tri-layer Ruddlesden-Popper nickelate $La_4Ni_3O_{10}$ under pressure and its experimental examination: Comparison with $La_3Ni_2O_7$. *Phys. Rev. B* **109**, 144511 (2024).

14. Wang, M., Wen, H.-H., Wu, T., Yao, D.-X. & Xiang, T. Normal and superconducting properties of $La_3Ni_2O_7$. *Chin. Phys. Lett.* (2024) doi:10.1088/0256-307X/41/7/077402.

15. Yang, J. et al. Orbital-dependent electron correlation in double-layer nickelate $La_3Ni_2O_7$. *Nat. Commun.* **15**, 4373 (2024).

16. Zhang, Y. et al. High-temperature superconductivity with zero resistance and strange-metal behavior in La$_3$Ni$_2$O$_{7-\delta}$. *Nat. Phys.* (2024) doi:10.1038/s41567-024-02515-y.

17. Dong, Z. et al. Visualization of oxygen vacancies and self-doped ligand holes in La$_3$Ni$_2$O$_{7-\delta}$. *Nature* **630**, 847–852 (2024).

18. Wang, G. et al. Pressure-Induced Superconductivity in Polycrystalline La$_3$Ni$_2$O$_{7-\delta}$. *Phys. Rev. X* **14**, 011040 (2024).

19. Puphal, P. et al. Unconventional Crystal Structure of the High-Pressure Superconductor La$_3$Ni$_2$O$_7$. *Phys. Rev. Lett.* **133**, 146002 (2024).

20. Zhou, Y. G. et al. Investigations of key issues on the reproducibility of high-Tc superconductivity emerging from compressed La$_3$Ni$_2$O$_{7-\delta}$. Preprint at arxiv.org/abs/2311.12361 (2023).

21. Chen, X. et al. Polymorphism in the Ruddlesden–Popper Nickelate La$_3$Ni$_2$O$_7$: Discovery of a Hidden Phase with Distinctive Layer Stacking. *J. Am. Chem. Soc.* **146**, 3640-3645 (2024).

22. Luo, Z., Hu, X., Wang, M., Wú, W. & Yao, D.-X. Bilayer Two-Orbital Model of La$_3$Ni$_2$O$_7$ under Pressure. *Phys. Rev. Lett.* **131**, 126001 (2023).

23. Christiansson, V., Petocchi, F. & Werner, P. Correlated Electronic Structure of La$_3$Ni$_2$O$_7$ under Pressure. *Phys. Rev. Lett.* **131**, 206501 (2023).

24. Liu, Y.-B., Mei, J.-W., Ye, F., Chen, W.-Q. & Yang, F. $s^{\pm}$-Wave Pairing and the Destructive Role of Apical-Oxygen Deficiencies in La$_3$Ni$_2$O$_7$ under Pressure. *Phys. Rev. Lett.* **131**, 236002 (2023).

25. Qu, X.-Z. et al. Bilayer Model and Magnetically Mediated Pairing in the Pressurized Nickelate. *Phys. Rev. Lett.* **132**, 036502 (2024).

26. Zhang, Y., Lin, L.-F., Moreo, A., Maier, T. A. & Dagotto, E. Structural phase transition, $s^{\pm}$-wave pairing, and magnetic stripe order in bi-layered superconductor La$_3$Ni$_2$O$_7$ under pressure. *Nat. Commun.* **15**, 2470 (2024).

27. Sreedhar, K. et al. Low-Temperature Electronic Properties of the La$_{n+1}$Ni$_n$O$_{3n+1}$ (n = 2, 3, and ∞) System: Evidence for a Crossover from Fluctuating-Valence to Fermi-Liquid-like Behavior. *J. Solid State Chem.* **110**, 208–215 (1994).

28. Zhang, Z., Greenblatt, M. & Goodenough, J. B. Synthesis, Structure, and Properties of the Layered Perovskite La$_3$Ni$_2$O$_{7-\delta}$. *J. Solid State Chem.* **108**, 402–409 (1994).

29. Chen, K. et al. Evidence of Spin Density Waves in La$_3$Ni$_2$O$_{7-\delta}$. *Phys. Rev. Lett.* **132**, 256503 (2024).

30. Fukamachi, T., Kobayashi, Y., Miyashita, T. & Sato, M. $^{139}$La NMR studies of layered perovskite systems La$_3$Ni$_2$O$_{7-\delta}$ and La$_4$Ni$_3$O$_{10}$. *J. Phys. Chem. Solids* **62**, 195–198 (2001).

31. Liu, Z. et al. Evidence for charge and spin density waves in single crystals of La$_3$Ni$_2$O$_7$ and La$_3$Ni$_2$O$_6$. *Sci. China Phys. Mech. Astron.* **66**, 217411 (2022).

32. Dan, Z. et al. Spin-density-wave transition in double-layer nickelate La$_3$Ni$_2$O$_7$. Preprint at https://doi.org/10.48550/arXiv.2402.03952 (2024).

33. Kobayashi, Y. et al. Transport and Magnetic Properties of La$_3$Ni$_2$O$_{7-\delta}$ and La$_4$Ni$_3$O$_{10-\delta}$. *J. Phys. Soc. Jpn.* **65**, 3978–3982 (1996).

34. Wang, L. et al. Structure responsible for the superconducting state in La$_3$Ni$_2$O$_7$ at high pressure and low temperature conditions. *J. Am. Chem. Soc.* **146**, 7506–7514 (2024).

35. Wang, G. et al. Observation of high-temperature superconductivity in the high-pressure tetragonal phase of La$_2$PrNi$_2$O$_{7-\delta}$. Preprint at https://doi.org/10.48550/arXiv.2311.08212 (2023).

36. Wang, N. et al. Bulk high-temperature superconductivity in pressurized tetragonal La$_2$PrNi$_2$O$_7$. *Nature* **634**, 579-584 (2024).

37. Khasanov, R. et al. Pressure-Induced Split of the Density Wave Transitions in La$_3$Ni$_2$O$_{7-\delta}$. Preprint at https://doi.org/10.48550/arXiv.2402.10485 (2024).

38. Zhang, J. et al. Intertwined density waves in a metallic nickelate. *Nat. Commun.* **11**, 6003 (2020).

39. Klotz, S., Chervin, J. C., Munsch, P. & Le Marchand, G. Hydrostatic limits of 11 pressure transmitting media. *J. Phys. D: Appl. Phys.* **42**, 075413 (2009).

40. Chen, X. et al. Electronic and magnetic excitations in La$_3$Ni$_2$O$_7$. *Nat. Commun.* **15**, 9597 (2024).

41. Wang Y., Jiang K., Wang Z., Zhang F.-C. and Hu J. Electronic and magnetic structures of bilayer La$_3$Ni$_2$O$_7$ at ambient pressure. *Phys. Rev. B* **110**, 205122 (2024)

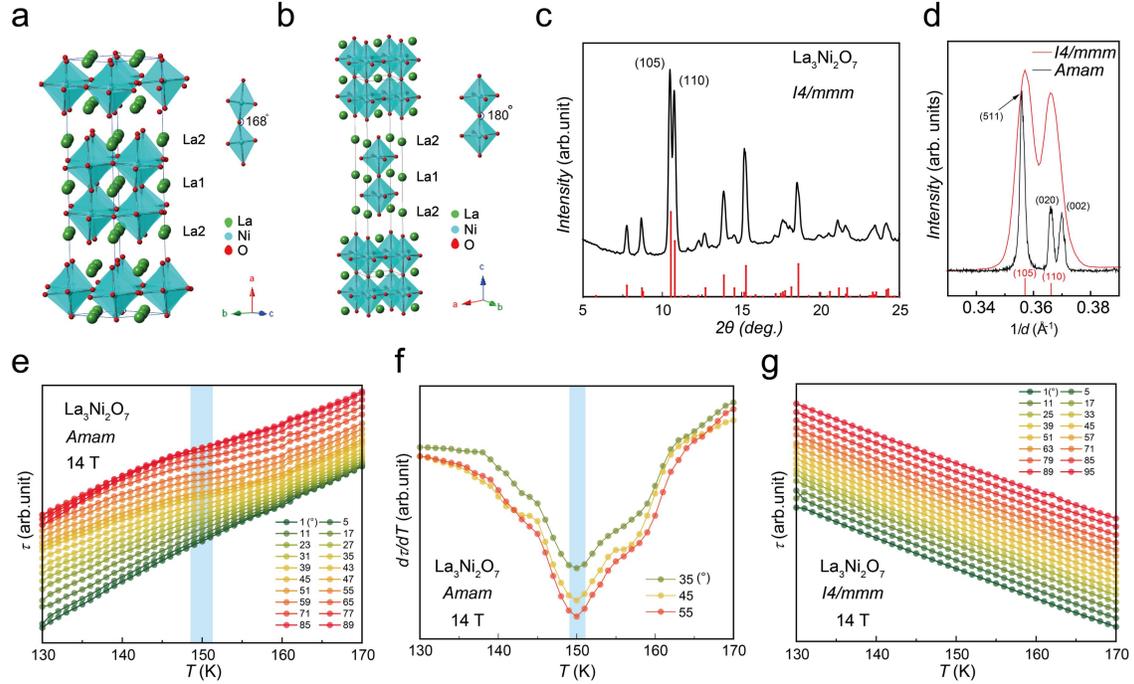

**Fig. 1. The structure and magnetic torque measurements for orthorhombic and tetragonal $La_3Ni_2O_{7-\delta}$ at ambient pressure.** (**a**) and (**b**): The crystal structure model of orthorhombic and tetragonal $La_3Ni_2O_{7-\delta}$ at ambient pressure, respectively. There is a tilting angel between the adjacent $NiO_6$ octahedrons for orthorhombic $La_3Ni_2O_{7-\delta}$, while the tilt along the c-axis disappears in the tetragonal $La_3Ni_2O_{7-\delta}$. (**c**): The powder XRD pattern for tetragonal $La_3Ni_2O_{7-\delta}$ at ambient pressure. (**d**): The enlarged XRD pattern with $1/d$ between 0.33-0.39 Å$^{-1}$, and the powder XRD pattern of the orthorhombic $La_3Ni_2O_{7-\delta}$ at ambient pressure is also potted for comparison. (**e**) and (**g**): Temperature-dependent magnetic torque data of orthorhombic and tetragonal $La_3Ni_2O_{7-\delta}$, respectively. (**f**): Temperature dependent $d\tau/dT$ at the angle of 35°, 45° and 55°. There is obvious feature in $\tau(T)$ around 150 K for orthorhombic $La_3Ni_2O_{7-\delta}$, implying the SDW transition which is consistent with previous μSR and NMR results. Such feature is not observed in the magnetic torque data for the tetragonal $La_3Ni_2O_{7-\delta}$, indicating no SDW transition for the tetragonal $La_3Ni_2O_{7-\delta}$.

**Table 1 Crystal data and structure refinement for orthorhombic and tetragonal $La_3Ni_2O_{7-\delta}$ at ambient pressure.**

| Identification code | $La_3Ni_2O_7$-OP | $La_3Ni_2O_7$-TP |
|---|---|---|
| Empirical formula | $La_3Ni_2O_{6.93}$ | $La_3Ni_2O_{6.96}$ |
| Formula weight | 645.07 | 645.43 |
| Temperature/K | 299.6(3) | 301.70(10) |
| Crystal system | orthorhombic | tetragonal |
| Space group | $Amam$ | $I4/mmm$ |
| a/Å | 20.5353(7) | 3.8542(2) |
| b/Å | 5.4479(2) | 3.8542(2) |
| c/Å | 5.3908(2) | 20.2302(18) |
| α/° | 90 | 90 |
| β/° | 90 | 90 |
| γ/° | 90 | 90 |
| Volume/Å$^3$ | 603.09(4) | 300.52(4) |
| Z | 4 | 2 |
| $\rho_{calc}$ g/cm$^3$ | 7.104 | 7.133 |
| μ/mm$^{-1}$ | 167.489 | 168.076 |
| F (000) | 1130.0 | 565.0 |
| Crystal size/mm$^3$ | 0.12 × 0.06 × 0.03 | 0.1 × 0.06 × 0.05 |
| Radiation | Cu Kα (λ = 1.54184) | Cu Kα (λ = 1.54184) |
| 2Θ range for data collection/° | 8.612 to 156.138 | 8.742 to 153.538 |
| Index ranges | -25 ≤ h ≤ 25, -3 ≤ k ≤ 6, -6 ≤ l ≤ 6 | -4 ≤ h ≤ 3, -4 ≤ k ≤ 4, -22 ≤ l ≤ 25 |
| Reflections collected | 3006 | 1341 |
| Independent reflections | 373 [Rint = 0.0630, Rsigma = 0.0299] | 128 [Rint = 0.0959, Rsigma = 0.0416] |
| Data/restraints/parameters | 373/0/31 | 128/0/17 |
| Goodness-of-fit on F$^2$ | 1.138 | 1.161 |
| Final R indexes [I>=2σ (I)] | R1 = 0.0374, wR2 = 0.1105 | R1 = 0.0677, wR2 = 0.1733 |
| Final R indexes [all data] | R1 = 0.0379, wR2 = 0.1110 | R1 = 0.0678, w2 = 0.1734 |
| Largest diff. peak/hole / e Å$^{-3}$ | 1.97/-2.17 | 4.51/-5.35 |

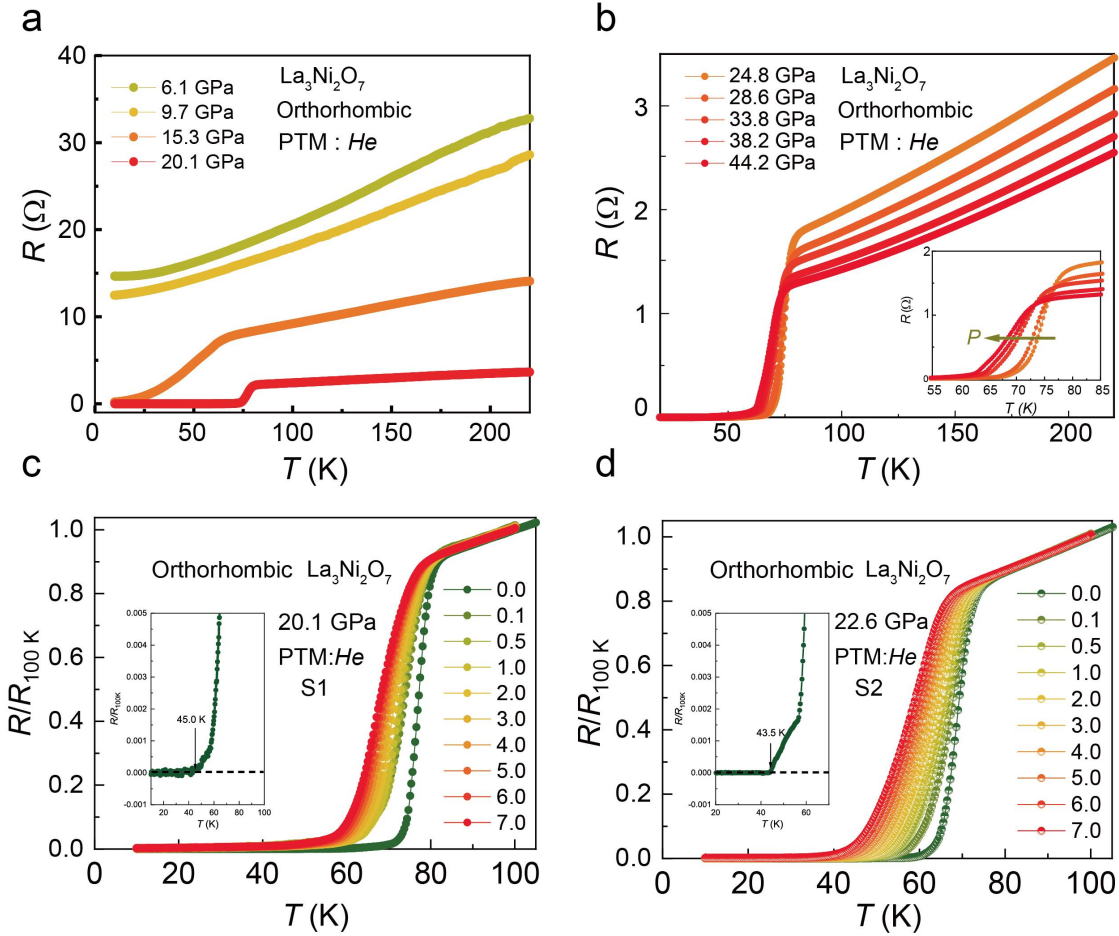

**Fig. 2. Temperature dependence of resistance (*R*) under various pressures for orthorhombic La₃Ni₂O₇₋δ.** (**a**) and (**b**): Temperature dependent resistance curves for the orthorhombic La₃Ni₂O₇₋δ under different pressures. The inset of (**b**) shows the gradual suppression of $T_c$ by pressure above 20 GPa. (**c**) and (**d**): The *R*(*T*) curves for the orthorhombic La₃Ni₂O₇₋δ under various magnetic field. The sample S1 and S2 are measured at 20.1 GPa and 22.6 GPa, respectively. The superconducting transition is gradually suppressed to a low temperature with increasing magnetic field. The insets clearly show that the temperature of zero resistance for the two single crystals of La₃Ni₂O₇₋δ is as high as 43-45 K, which is highest temperature of zero resistance reported in the La₃Ni₂O₇₋δ single crystals so far.

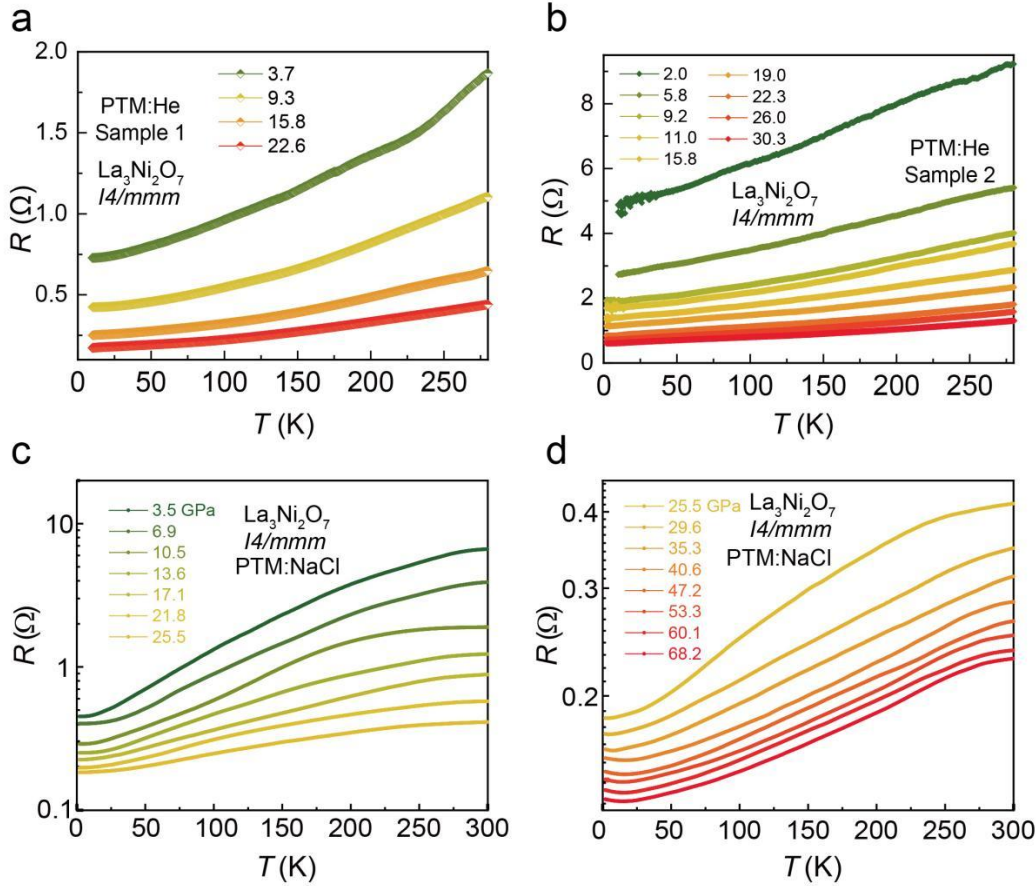

**Fig. 3. Temperature dependence of resistance ($R$) under various pressures for tetragonal La$_3$Ni$_2$O$_{7-\delta}$.** (**a**) and (**b**): Temperature dependent resistance for tetragonal La$_3$Ni$_2$O$_{7-\delta}$ under various pressure for two samples. The $R(T)$ curves show a metallic behavior without any trace originated from superconducting transition or SDW under the pressure up to 30.3 GPa in the whole temperature range. All the above measurements were conducted on a DAC using Helium gas as the pressure transmitting medium so as to get a more homogeneous pressure environment. (**c**) and (**d**): The $R(T)$ curves for tetragonal La$_3$Ni$_2$O$_{7-\delta}$ at various pressures using NaCl as the pressure transmitting medium. The resistance continually decreases with increasing the pressure, and no anomaly arisen from the superconductivity or SDW is observed. To compare the two different pressure transmitting media, $R(T)$ shows some difference due to pressure inhomogeneity, especially a slight upturn shows up at low temperature below 20 K under high pressure above 40 GPa for the case of using NaCl as the pressure transmitting medium.

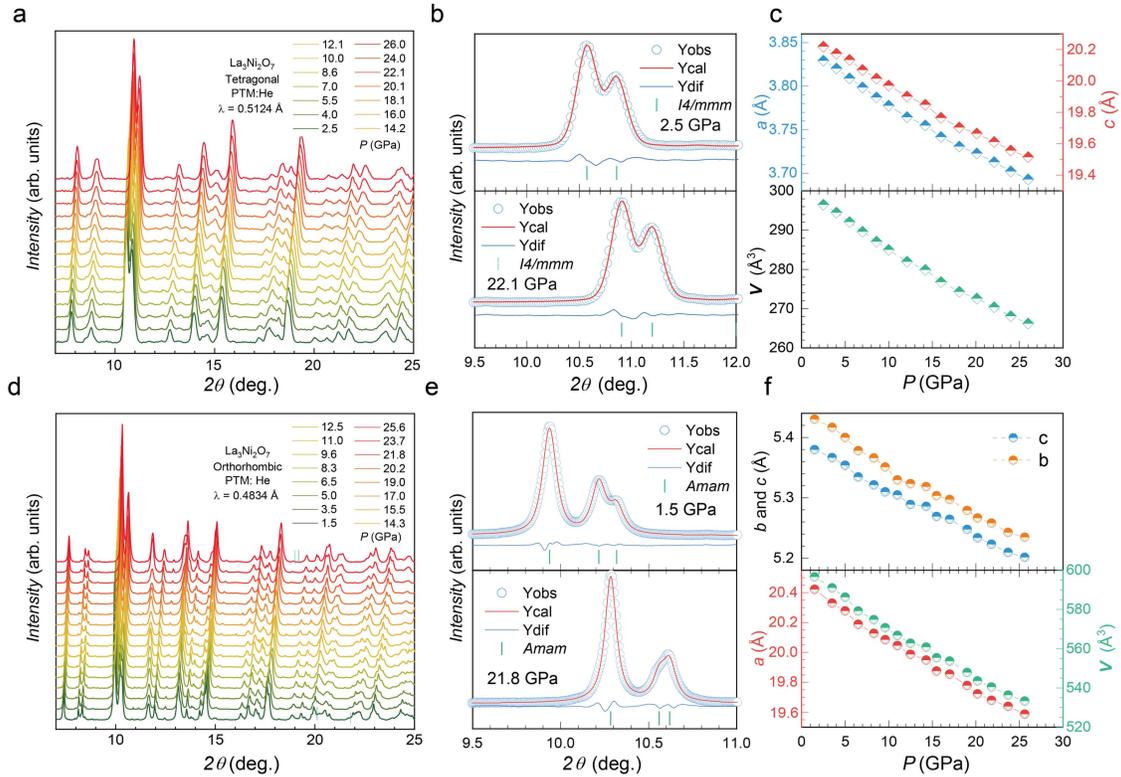

**Fig. 4. Evolution of X-ray diffraction patterns and lattice parameters with pressure for the tetragonal and orthorhombic $La_3Ni_2O_{7-\delta}$.** (**a**): The powder XRD patterns for the tetragonal $La_3Ni_2O_{7-\delta}$ under different pressures. (**b**): Rietveld refinement of powder XRD patterns for the tetragonal $La_3Ni_2O_{7-\delta}$ under 2.5 GPa (the upper panel) and 22.1 GPa (the lower panel), respectively. The blue circles and red lines represent the observed and calculated data, respectively. The blue lines indicate the difference between the observed and calculated data. The short green vertical lines indicate the calculated diffraction peak positions. The powder XRD patterns can be well fitted with the space-group *I4/mmm* at 2.5 GPa and 22.1 GPa. (**c**): The evolution of lattice parameters and cell volume with pressure for the tetragonal $La_3Ni_2O_{7-\delta}$. The blue and red dots are the lattice parameters for *a*-axis and *c*-axis, respectively. (**d**): The powder XRD patterns for the orthorhombic $La_3Ni_2O_{7-\delta}$ under different pressures. (**e**): Rietveld refinement of powder XRD patterns for the orthorhombic $La_3Ni_2O_{7-\delta}$ under 1.5 GPa (the upper panel) and 21.8 GPa (the lower panel) using the space-group of *Amam*. The Extended Data Fig. 5 shows the Rietveld refinement of powder XRD pattern taken at 21.8 GPa using *Amam* and *Fmmm* as the space group. Both the *Amam* and *Fmmm* space group fit well with the measured data, and we cannot determine its space group based on the current data. (**f**): The evolution of lattice parameters and cell volume with pressure for the orthorhombic $La_3Ni_2O_{7-\delta}$. The blue, orange and red dots are the lattice parameters for *c*-axis, *b*-axis and *a*-axis, respectively. Helium gas was used as the pressure transmitting medium during the measurement of all these powder XRD patterns so as to get a more homogeneous pressure environment.

**Method**

**Sample growth:** The $La_3Ni_2O_{7-\delta}$ crystals are grown through a melt salt method. Firstly, the Lanthanum nitrate hexahydrate, nickel (II) nitrate hexahydrate and citric acid (CA) were dissolved in the water with the mole ratio CA: La: Ni = 5:3:2. After preheated at 140 °C in an oven for about 24 h, the above product was transferred into a muffle furnace where the temperature is slowly increased to 400 °C in 10 h and kept for another 10 h. The above precursor (P) is mixed with a salt flux (NaCl/KCl mixture) at the mass ratio of P: NaCl: KCl = 1:14:16 and loaded into a corundum crucible. The corundum crucible is heated in the air to 1150 °C in 10 h and kept for 48 h, and then slowly cooled down to 1110 °C in 7 days. Micro-crystal with typical size 0.1×0.1×0.03 mm was obtained after washing the flux using water. The as grown product is orthorhombic $La_3Ni_2O_{7-\delta}$ with an insulating behavior due to oxygen deficiency. Then the micro-crystals are annealed at high oxygen pressure to obtain metallic $La_3Ni_2O_{7-\delta}$. The annealing temperature is slowly increased to 500 °C in 10 h and kept for 5 h, then slowly cooled down to 50 °C in 6 days. For metallic orthorhombic $La_3Ni_2O_{7-\delta}$, the sample is annealed at 10 - 15 bar oxygen. For metallic tetragonal $La_3Ni_2O_{7-\delta}$, the sample is annealed at 150 bar oxygen. The refined oxygen content for metallic sample based on the single crystal X-ray diffraction (SC-XRD) data is $La_3Ni_2O_{6.93}$ for the orthorhombic phase and $La_3Ni_2O_{6.96}$ for the tetragonal phase, respectively.

**Ambient-pressure structural and magnetic torque characterization:** The single crystal X-ray diffraction (SC-XRD) data was collected on a four-circle diffractometer (Rigaku, XtaLAB PRO 007HF) with Cu K$a$ radiation in Core Facility Center for Life Sciences, USTC. The structure was solved and refined using Olex-2 with ShelXT and ShelXL packages. The detailed structure data is shown in Table 1. Torque magnetometry was conducted in a Physical Properties Measurement System (PPMS, Quantum Design Inc., DynaCool-14T) using an SCL piezoresistive cantilever. The sample was attached to the tip of cantilever which was fixed on a horizontal rotator. We first rotated the sample in a range of $\theta$ (the angle between magnetic field vector $H$ and the flat plane of $La_3Ni_2O_{7-\delta}$ crystal) from 0° to 90° under isothermal condition and determined that the largest signal occurs at approximately $\theta$ close to 45°.

**High pressure electrical transport and XRD measurements:** The high-pressure resistance for $La_3Ni_2O_{7-\delta}$ single crystals were measured in the diamond anvil cells by using NaCl or Helium gas as the pressure transmitting medium. Diamond anvils with various culets (200 to 400 μm) were used for high-pressure transport measurements. The pressure was applied and calibrated by the shift of ruby florescence at room temperature. The transport measurements were carried out in a refrigerator system (HelioxVT, Oxford Instruments) or Physical Properties Measurement System (PPMS-9, Quantum Design Inc.). The powder XRD data of orthorhombic $La_3Ni_2O_{7-\delta}$ under pressure is collected at Shanghai at Shanghai Synchrotron Radiation Facility, using an X-ray beam with a wavelength of 0.4834 Å. The Helium gas or silicon oil was used as the pressure transmitting medium for orthorhombic $La_3Ni_2O_{7-\delta}$. The powder XRD data of tetragonal $La_3Ni_2O_{7-\delta}$ under pressure is collected using a MetalJet E1+ 160 kV source equipped with an In-Ga target, generating an X-ray wavelength of 0.5124 Å. The Helium gas was used as the pressure transmitting medium for tetragonal $La_3Ni_2O_{7-\delta}$. Two-dimensional diffraction images were recorded by a PILATUS R CdTe detector and subsequently processed into one-dimensional XRD patterns using the Dioptas software. The powder XRD data were refined using the Jana2020 or GSAS-II software to obtain the lattice parameters under different pressures.

## Data availability

The data that support the findings of this study are available from the corresponding author upon request.

## Code availability

The code that supports the findings of this study are available from the corresponding author upon request.

## Acknowledgments

We acknowledge fruitful discussions with Ho-kwang Mao, Zhengyu Wang and Ziji Xiang. We also thank Zhongliang Zhu, Fujun Lan, Yuxin Liu, and Hongbo Lou for their experiment assistance. This work is supported by the National Natural Science Foundation of China (Grants No. 12494592, 12488201, 11888101, 12034004, 12161160316, 12325403), the National Key R&D Program of the MOST of China (Grant No. 2022YFA1602601), the Strategic Priority Research Program of Chinese Academy of Sciences (Grant No. XDB25000000), the Chinese Academy of Sciences under contract No. JZHKYPT-2021-08, the CAS Project for Young Scientists in Basic Research (Grant No. YBR-048), and the Innovation Program for Quantum Science and Technology (Grant No. 2021ZD0302800). D.P. and Q.Z. acknowledge the financial support from the Shanghai Science and Technology Committee (Grant No. 22JC1410300) and Shanghai Key Laboratory of Material Frontiers Research in Extreme Environments (Grant No.22dz2260800). A portion of this research used resources at the beamline 17UM of Shanghai synchrotron radiation facility.

## Author contributions

X.H.C. conceived the research project and coordinated the experiments. M.Z.S. grew the single crystals and performed the structural characterization at ambient pressure; H.P. L. and K.B. F. measured the magnetic torque data; Y.K.L., R.Q. W. and J.J.Y. performed the high-pressure transport measurement using NaCl as the pressure-transmitting medium; D.P. performed the resistance measurements using helium as a pressure-transmitting medium under pressure with the help of Q.S.Z.; D.P., Z.F.X. and Y.Z.W. performed the synchrotron powder diffraction measurements and analysis under high-pressure using helium and silicon oil as the pressure-transmitting medium with help from Q.S.Z. and Z.D.Z.; M.Z.S., D.P., J.J.Y., and X.H.C. analyzed the data; J.J.Y., M.Z.S., D.P., T.W. and X.H.C. wrote the paper with inputs from all authors.

# Extended Figures and Tables

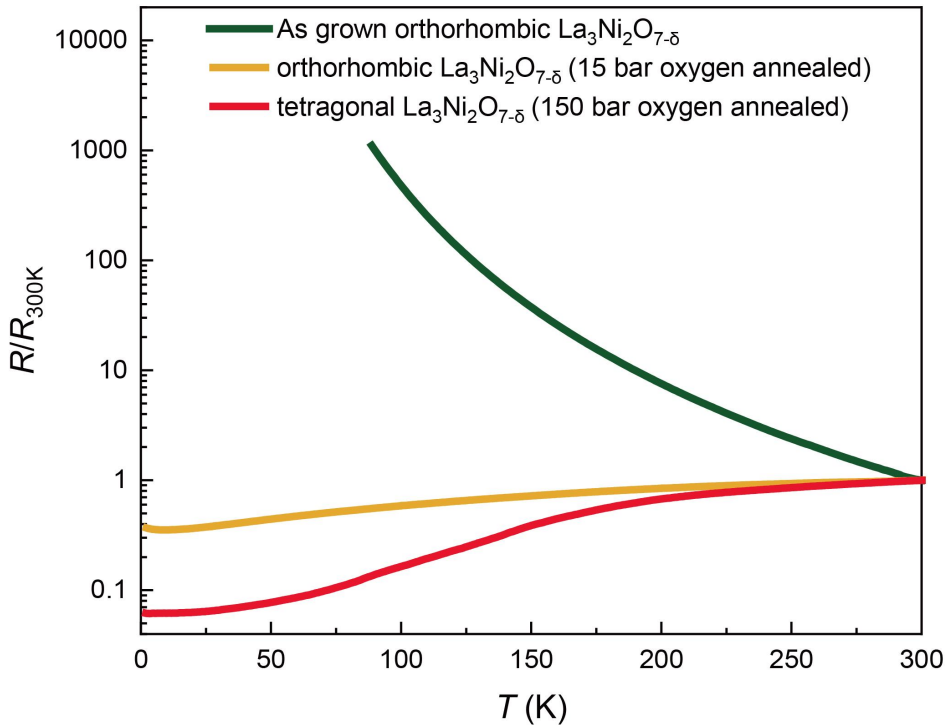

**Extended data Fig. 1. Temperature dependent normalized resistance curves for the as grown and oxygen annealed $La_3Ni_2O_{7-\delta}$ samples.** The resistance measurement on the powder samples for the orthorhombic $La_3Ni_2O_{7-\delta}$ as grown samples and orthorhombic $La_3Ni_2O_{7-\delta}$ samples annealed at oxygen pressure of 15 bar were conducted, while the resistance measurement on microcrystal for the tetragonal $La_3Ni_2O_{7-\delta}$ annealed at oxygen pressure of 150 bar was conducted under near ambient pressure (0.9 GPa).

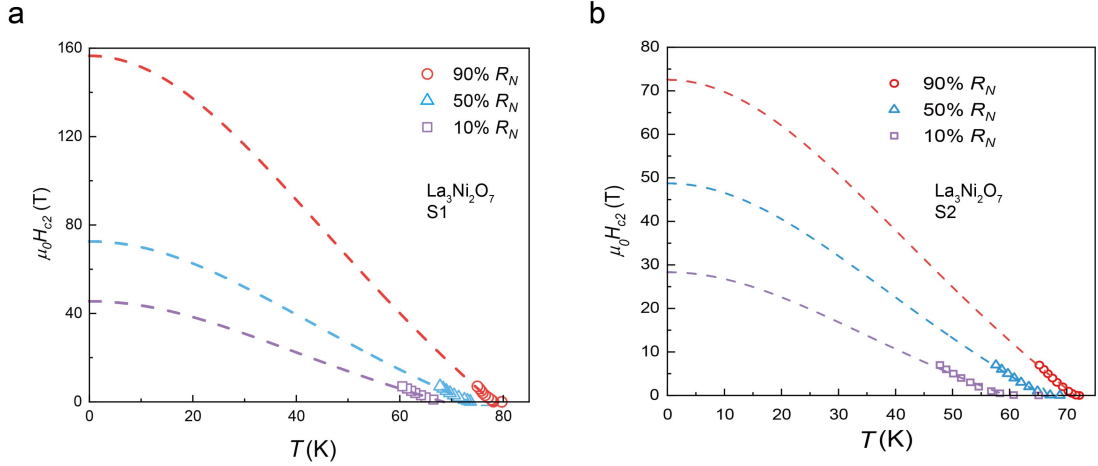

**Extended Data Fig. 2. The upper critical field for two orthorhombic $La_3Ni_2O_{7-\delta}$ samples.** The upper critical fields were fitted with different criteria using a 3D GL model. In the 3D GL model, the $H_c$-$T$ relation for a superconductor follows the equation

$$\mu_0 H_{c2}(T) = \mu_0 H_{c2}(0) \frac{1-(\frac{T}{T_C})^2}{1+(\frac{T}{T_C})^2},$$

where $H_{c2}$ and $T_c$ are the upper critical field and superconducting temperature.

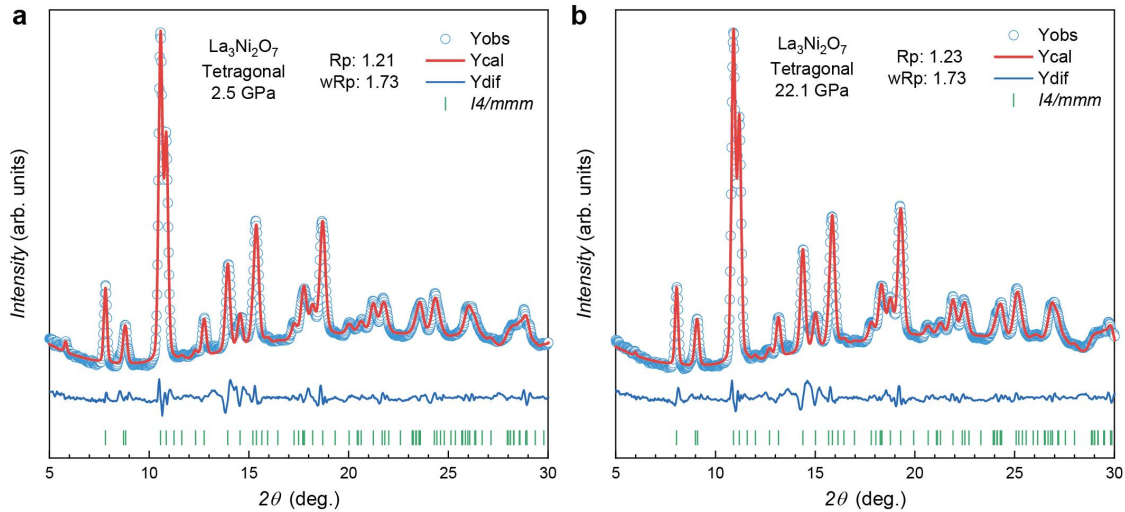

**Extended Data Fig. 3. Rietveld refinement of powder XRD data for the tetragonal $La_3Ni_2O_{7-\delta}$.** The refined powder XRD patterns under 2.5 GPa (**a**) and 22.1 GPa (**b**), respectively. The blue circles and red lines represent the observed and calculated data, respectively. The blue lines indicate the difference between the observed and calculated data. The short green vertical lines indicate the calculated diffraction peaks positions. The powder XRD data can be well fitted with the space-group *I4/mmm* under both 2.5 GPa and 22.1 GPa, which indicates no structure transitions under pressure up to 26 GPa for the tetragonal $La_3Ni_2O_{7-\delta}$.

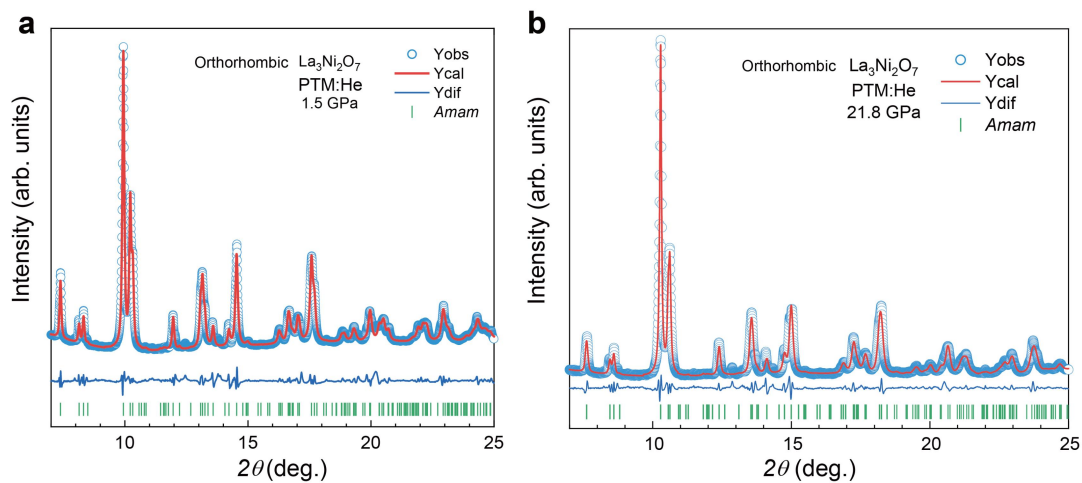

**Extended Data Fig. 4. Rietveld refinement of the powder XRD patterns for the orthorhombic La$_3$Ni$_2$O$_{7-\delta}$.** The refined powder XRD data under 1.5 GPa (**a**) and 21.8 GPa (**b**), respectively. The blue circles and red lines represent the observed and calculated data, respectively. The blue lines indicate the difference between the observed and calculated data. The short green vertical lines indicate the calculated diffraction peaks positions. The XRD data can be well fitted with the space group *Amam* for the orthorhombic La$_3$Ni$_2$O$_{7-\delta}$ at 1.5 GPa and 21.8 GPa, respectively.

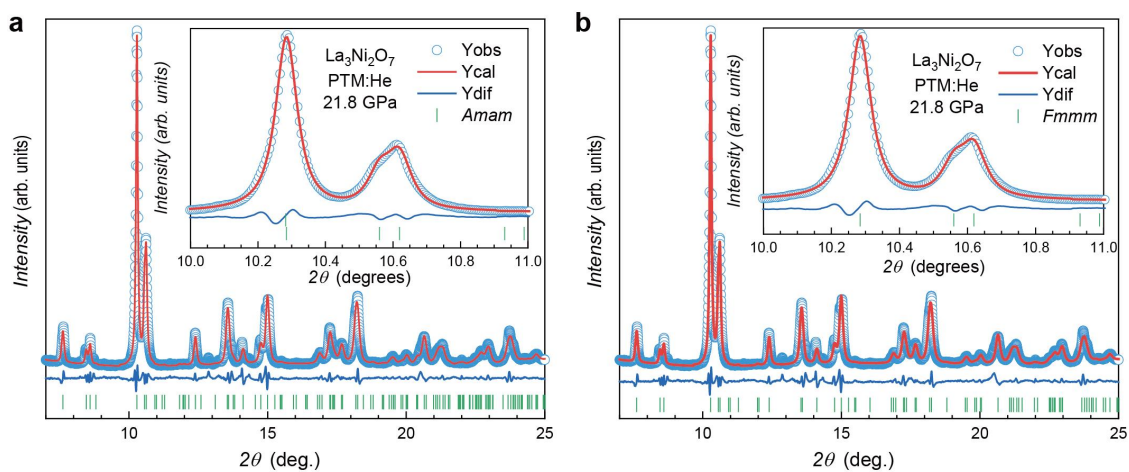

**Extended Data Fig. 5. Rietveld refinement of the powder XRD patterns for the orthorhombic La$_3$Ni$_2$O$_{7-\delta}$.** The refined powder XRD patterns under 21.8 GPa using the space group of *Amam* (**a**) and *Fmmm* (**b**), respectively. The inset shows the enlarged view between 10-11°. The blue circles and red lines represent the observed and calculated data, respectively. The blue lines indicate the difference between the observed and calculated data. The short green vertical lines indicate the calculated diffraction peak positions. The powder XRD data under 21.8 GPa can be well fitted with the space group *Amam* and *Fmmm*, respectively. It should be addressed that we cannot determine its space group based on the current data.

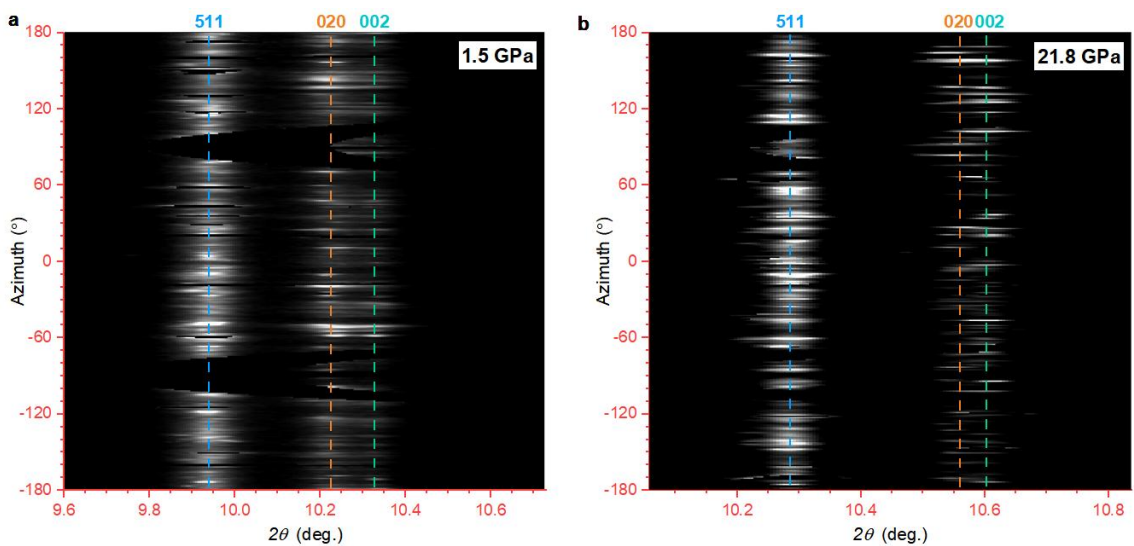

**Extended Data Fig. 6. Cake representation of the typical X-ray powder diffraction pattern measured for the orthorhombic La₃Ni₂O₇₋δ.** The cake data of the area detector measured at 1.5 GPa (a) and 21.8 GPa (b), respectively. It further indicates no structural transition from orthorhombic to tetragonal with increasing pressure up to 21.8 Gpa.

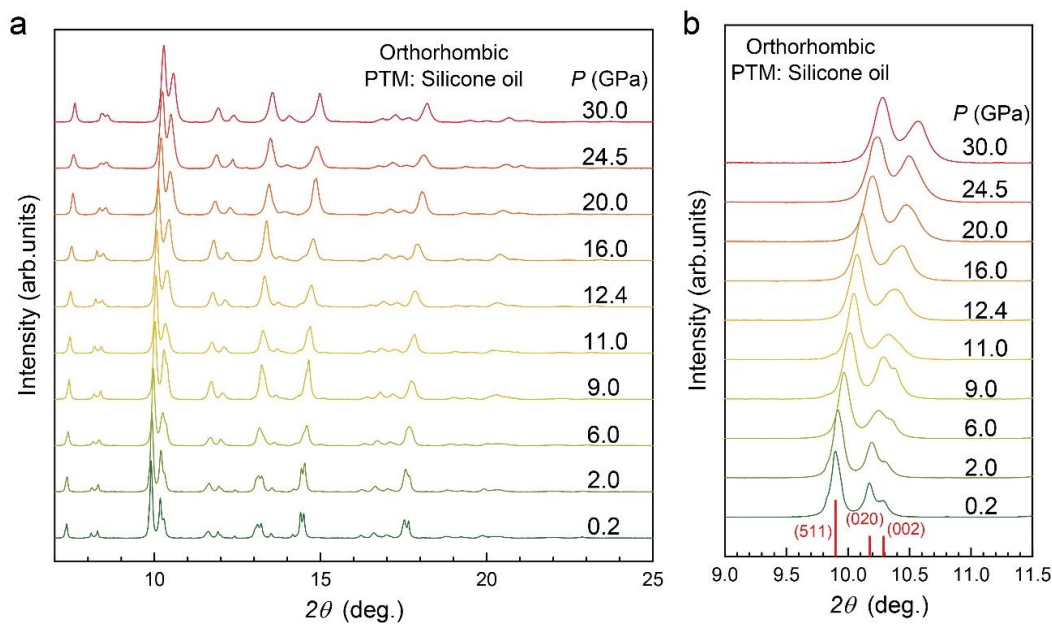

**Extended Data Fig. 7. Structural evolution of the orthorhombic La₃Ni₂O₇₋δ under various pressures using silicon oil as the pressure transmitting medium.** (**a**): the powder XRD patterns under different pressures for the orthorhombic La₃Ni₂O₇₋δ using silicon oil as the pressure transmitting medium. (**b**): The enlarged view of (**a**) with 2$\Theta$ between 9-11.5°. The diffraction peaks (020) and (002) gradually merges together with increasing pressure above 11 GPa.